\documentclass[]{spie}
\usepackage{graphicx}
\DeclareGraphicsExtensions{.pdf,.png,.jpg}
\usepackage[caption=false
]{subfig}
\usepackage{lipsum}
\usepackage{wrapfig}
\usepackage{booktabs}
\usepackage{lmodern}
\graphicspath{{figures/}}

\title{A patient-specific scatter artifacts correction method }

\author{Wei Zhao\supit{a}, Stephen Brunner\supit{a}, Kai Niu\supit{a}, Sebastian Schafer\supit{b}, Kevin Royalty\supit{b},\\Guang-Hong Chen\supit{a,c}
\skiplinehalf
\supit{a}Department of Medical Physics, University of Wisconsin-Madison, WI 53705;\\
\supit{b}Siemens Medical Solutions, USA, Inc., Hoffman Estates, IL 60192;\\
\supit{c}Department of Radiology, University of Wisconsin-Madison, WI 53792;
}

\begin{document}

\maketitle

\begin{abstract}
This paper provides a fast and patient-specific scatter artifact correction method for cone-beam computed tomography (CBCT) used in image-guided interventional procedures.  Due to increased irradiated volume of interest in CBCT imaging, scatter radiation has increased dramatically compared to 2D imaging, leading to a degradation of image quality. In this study, we propose a scatter artifact correction strategy using an analytical convolution-based model whose free parameters are estimated using a rough estimation of scatter profiles from the acquired cone-beam projections. It was evaluated using Monte Carlo simulations with both monochromatic and polychromatic X-ray sources. The results demonstrated that the proposed method significantly reduced the scatter-induced shading artifacts and recovered CT numbers.
\end{abstract}


\section{Introduction}
Due to increased irradiated image volumes, scatter radiation has been one of the major challenges in flat-panel detector based cone-beam computed tomography (CBCT). Typical visual appearances of scatter artifacts include cupping, dark streaks between dense objects, degraded contrast resolution, and inaccurate CT numbers for quantitative measurements. Many scatter correction methods have been proposed since the early days of x-ray imaging. These include scatter rejection techniques\cite{Ruhrnschopf2011} such as the use of an air gap, antiscatter grid and bow-tie filter, analytical modeling of scatter profiles\cite{Boone1988,Seibert1988,Ohnesorge1999,Li2008,Maltz2008,Star-Lack2009,Sun2010,Meyer2010a,Star-Lack2013}, Monte Carlo simulations\cite{Kyriakou2006,Jarry2006,Bertram2008,Malusek2008,Colijn2004,Zbijewski2006,Mainegra-Hing2010,Poludniowski2009}, primary modulation methods using either a spatial primary modulator\cite{Maltz2006,Zhu2006} or a temporal primary modulator\cite{Schorner2012} and other measurement-based scatter correction methods which measure scatter radiation\cite{Ning2004,Zhu2005,Siewerdsen2006,Liu2006,Zhu2009,Jin2010,Yan2010,Wang2010,Lee2012,Niu2011,Ouyang2013} using either a beam stop\cite{Ning2004}, an aperture\cite{Siewerdsen2006}, or a moving blocker\cite{Zhu2005,Wang2010,Ouyang2013}.

While these methods have been demonstrated to mitigate scatter artifacts, some challenges still remain. For example, scatter rejection techniques usually yield inadequate correction and additional scatter corrections are needed \cite{Ruhrnschopf2011}. Scatter rejection techniques can also reduce soft-tissue contrast-to-noise ratio (CNR) due to the concomitant attenuation of primary radiations\cite{Schafer2012}. In current analytical modeling method, both calibration measurements and Monte Carlo (MC) simulations may be needed to calibrate the free parameters in the analytical modeling of scatter profiles. MC simulations can be computationally expensive and may take prohibitively long processing time.  In primary modulation methods, the necessary extra hardware support or mechanical modifications add complexities to the current image acquisition system and thus become difficult to adapt into a clinical setting. In direct scatter measurement methods,  extra scans or extra hardware or a dedicated reconstruction algorithm may be needed. As a result, it is highly desirable to develop a scatter artifacts correction method that is fast, patient specific, sufficiently accurate and, ideally, does not require additional scans, or extra hardware support and modifications. In this paper, we present a new method to achieve this goal.

\section{Methods}

The flowchart of the proposed algorithm is presented in Figure 1.  The proposed algorithm estimates scatter radiation distribution using a coarse scatter distribution which is regularized with a convolution-based scatter model. The coarse scatter estimation is achieved by taking the difference between the raw projection data and the polychromatic reprojection data of a segmented image volume. The method includes three major components: convolution-based scatter estimation model, coarse scatter estimation and calibration of free parameters used in the convolution-based model.


\begin{figure}[t]
    \centering
    \includegraphics[width=0.8\textwidth]{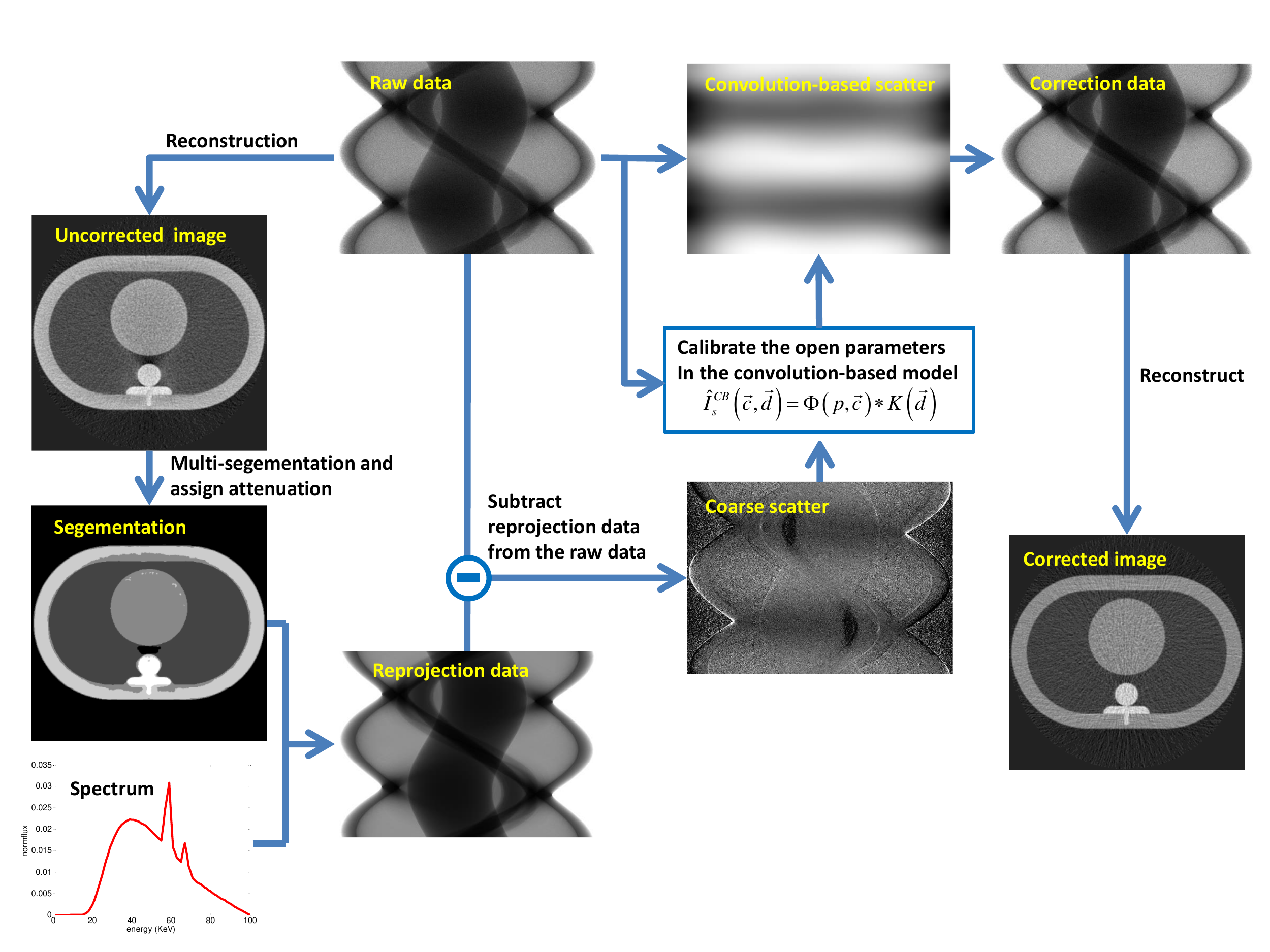}
    \vspace{-1em}
    \caption{Flowchart of the proposed scatter correction method.}
    \vspace{-1em}
    \label{fig:f1}
\end{figure}

In the convolution-based scatter model, scatter signal $\hat{I}^{CB}_{s}$ can be expressed as a convolution between a scatter potential $\Phi$ and a scatter kernel $K$\cite{Baer2012}:
\begin{equation}
\label{ConvolutionModel}
\hat{I}^{CB}_{s}(\alpha,\vec{c},\vec{d}) = \Phi[p(\alpha),c_{0},c_{1}]\ast K(d_{1},d_{2}),
\end{equation}
where $\Phi[p(\alpha),\vec{c}] = c_{0}+c_{1}\cdot pe^{-p}$ and $K(\vec{d}) = e^{-d_{1}(x+d_{2})^2}+e^{-d_{1}(x-d_{2})^2}$ with $\vec{c}=(c_{0},c_{1})$ and $\vec{d}=(d_{1},d_{2})$ respectively. Here $p$ denotes the raw projection and $x$ denotes the detector index. The constant $c_{0}$ represents the contributions from the Compton scattering while the second term was attributed to the Rayleigh scattering\cite{Baer2012}. Although the scatter radiation distribution can be estimated using the above convolution-based model effectively, a calibration experiment is often needed to determine the free parameters sets $\vec{c},\vec{d}$. Methods have been proposed to determine these free parameters using either Monte Carlo simulations or a dedicated calibration scan. In this paper, we employed a novel  coarse scatter estimation method to determine these free parameters.


The raw projection data $I$ was modeled as a summation of primary beam $I_{p}$ and  scatter contribution $I_{s}$,
\begin{equation}\label{equ:est}
I = I_{p}+I_{s}
\end{equation}
Instead of directly estimate $I_{s}$, we generated an estimate of the primary signal  $\hat{I}_{p}$ which was estimated to be the forward projection of segmented image volume using a polychromatic x-ray spectrum.  To estimate the scatter contribution, we simply subtracted the estimated primary signal $\hat{I}_{p}$ from the measured raw projection data $I$ to obtain the needed coarse estimation of the scatter signal  $\hat{I}_{s}$.

The estimated scatter profile $\hat{I}_{s}$ was used in  the analytical convolution model in Eq. (\ref{ConvolutionModel}) to calibrate the unknown parameters $\vec{c}$ and $\vec{d}$. By minimizing the least square error between $\hat{I}_{s}$ and $\hat{I}^{CB}_{s}$, the unknown parameters could be obtained. Using the obtained final analytical convolution model, the final scatter radiation distribution was calculated and the corrected projection data was then obtained by subtracting the final scatter signal from the raw projection data. Using the scatter corrected projection data, the routine FBP reconstruction method was used to obtain CT images with reduced scatter induced artifacts.

\section{Numerical Simulations}

In this study, an anthropomorphic quasi-thorax phantom\footnote{http://www.qrm.de/content/pdf/QRM-Thorax.pdf.} was used to generate both primary beam and scatters using a Geant4 based Monte Carlo simulation package, GATE\cite{Jan2011}. Our numerical simulation studies contained two aspects: in the first numerical simulation study, a monochromatic x-ray beam at $70~keV$ was used since beam hardening artifacts have similar profiles as scatter artifacts. To be realistic, a polychromatic x-ray spectrum at $100~kVp$ was used in our second numerical simulation study. These two studies were evaluated to demonstrate the proposed method, since the polychromatic forward projection model can be used to take the beam hardening contribution in account during the coarse scatter estimation. The fan angle and the cone angle of the simulated data acquisitions are $23.6^{\circ}$ and $16^{\circ}$, respectively. The geometrical configuration of the simulated CBCT system was similar to  the Varian 2100 EX System used in image-guided radiation therapy. Scatter projection, primary projection, and total projection were recorded separately. During the correction procedure, the Monte Carlo simulation toolkit Geant4 was applied to generate the polychromatic X-ray spectrum for the $100~kVp$ case.




\section{Results}
\subsection{Monochromatic case}

\begin{figure}
    \centering
    \includegraphics[width=0.8\textwidth]{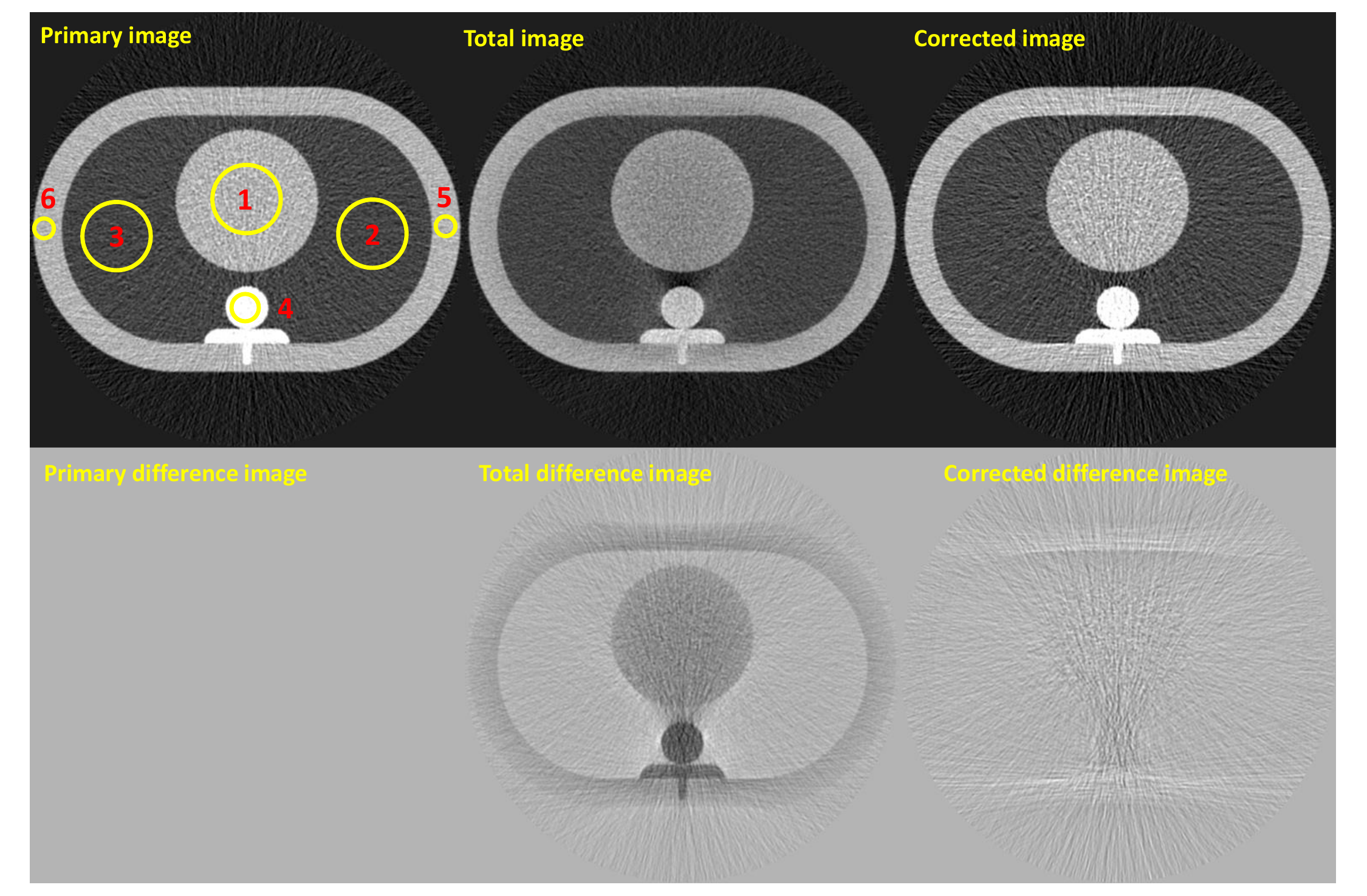}
    \caption{Scatter correction for the quasi-thorax phantom with $70~keV$ monochromatic X-ray source. After correction, scatter artifacts are almost completely removed and CT number are comparable to the primary image. The difference images show each image minus the scatter-free primary image. Display window: [-1200HU, 500HU] for both the CT images and the difference images. }
    \label{fig:f2}
\end{figure}

\begin{figure}
    \centering
    \includegraphics[width=\textwidth]{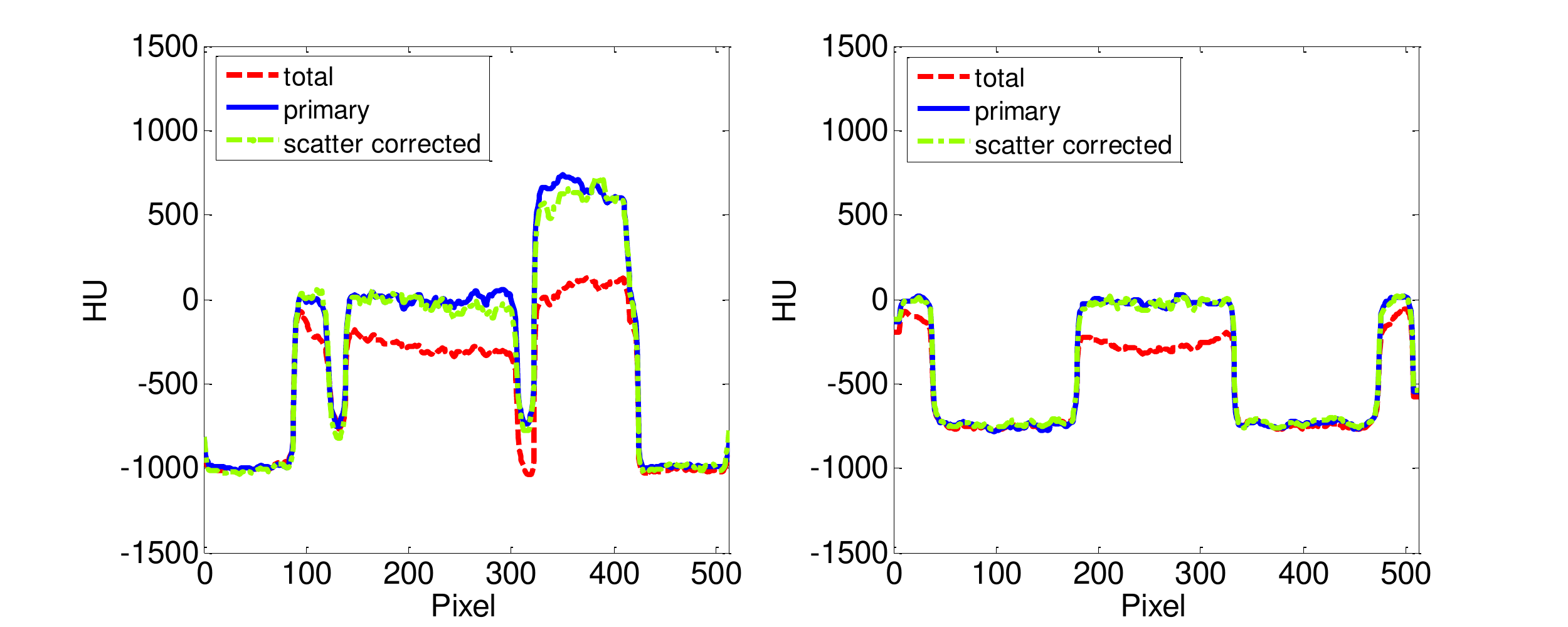}
    \caption{Line profiles of the quasi-thorax phantom with monochromatic $70~keV$ X-ray source. Shading and reduced CT number caused by scatter radiation are clearly shown by the red line. After correction with the proposed method, shading artifacts are well removed and CT number are well recovered, and the corrected profiles match the scatter-free primary profiles well. (a) Vertical direction, (b) horizontal direction. }
    \label{fig:f3}
\end{figure}

Fig.~\ref{fig:f2} shows the results of a scatter-free primary image, an image with both primary and scatter contributions, and the corrected CT image using the proposed algorithm. To help visualize the performance of the algorithm, the difference between the aforementioned three images and the scatter-free primary image was calculated and shown in the second row in  Fig.~\ref{fig:f2} .  In the corrected image, scatter artifacts are accurately removed and the CT number is restored to be closer to the primary image. The average errors in CT number were reduced from -190 HU to -11 HU for the different ROIs (shown in Figure.~\ref{fig:f2} upper-left corner). Fig.~\ref{fig:f3} shows the line profiles of the quasi-thorax phantom along central vertical direction and central horizontal direction (shown in Figure.~\ref{fig:f4} upper-left corner). Scatter caused shading artifacts are almost completely removed and the reduced CT number are successfully recovered. The corrected profiles match the scatter-free primary profiles very closely.

\subsection{Polychromatic case}

\begin{figure}
    \centering
    \includegraphics[width=0.8\textwidth]{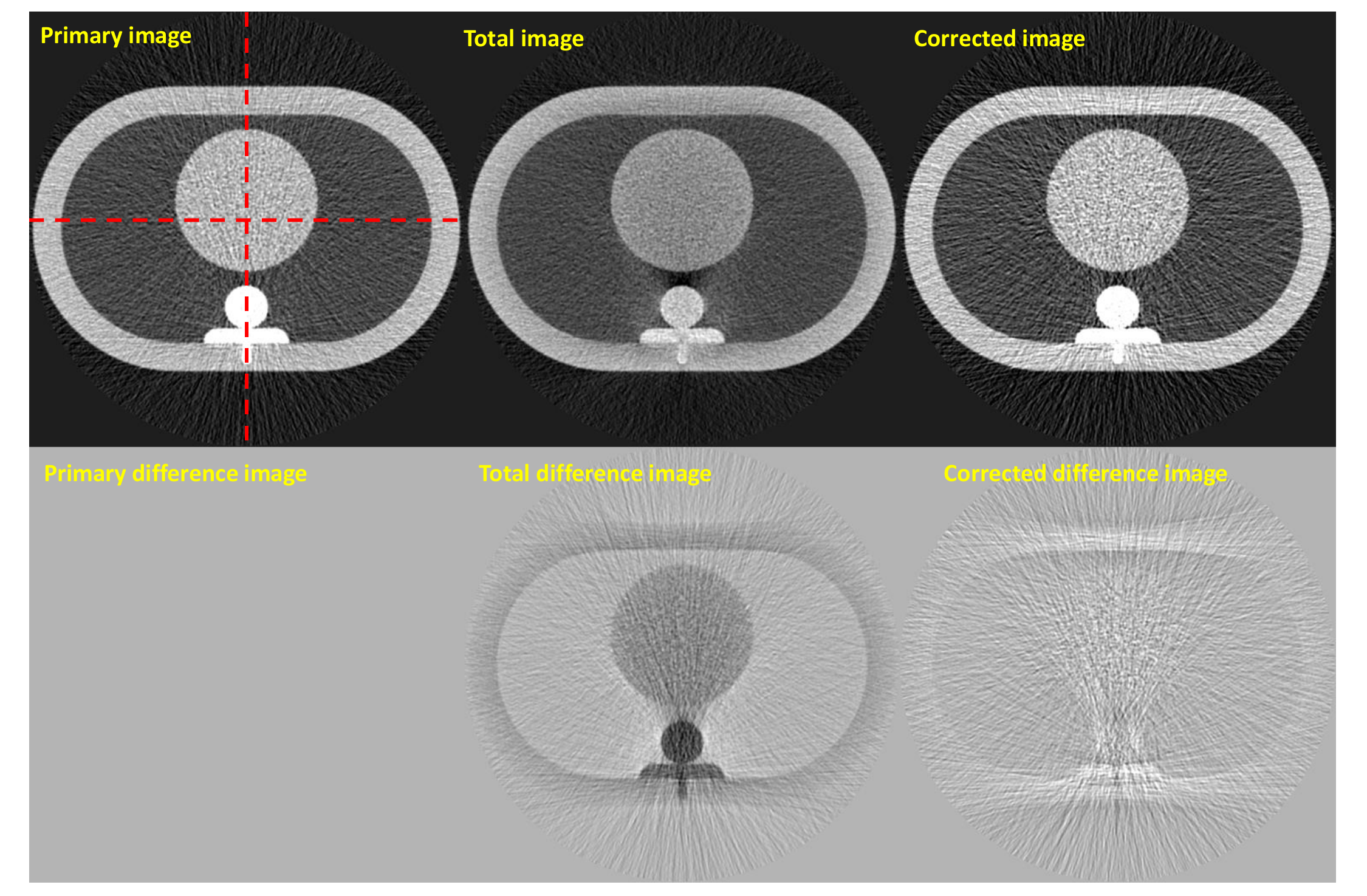}
    \caption{Scatter correction for the quasi-thorax phantom with $100~kVp$ polychromatic X-ray source. After correction, scatter artifacts are almost completely removed and CT number are comparable to the primary image. The difference images show each image minus the scatter-free primary image. Display window: [-1200HU, 500HU] for both the CT images and the difference images. }
    \label{fig:f4}
\end{figure}

\begin{figure}
    \centering
    \includegraphics[width=\textwidth]{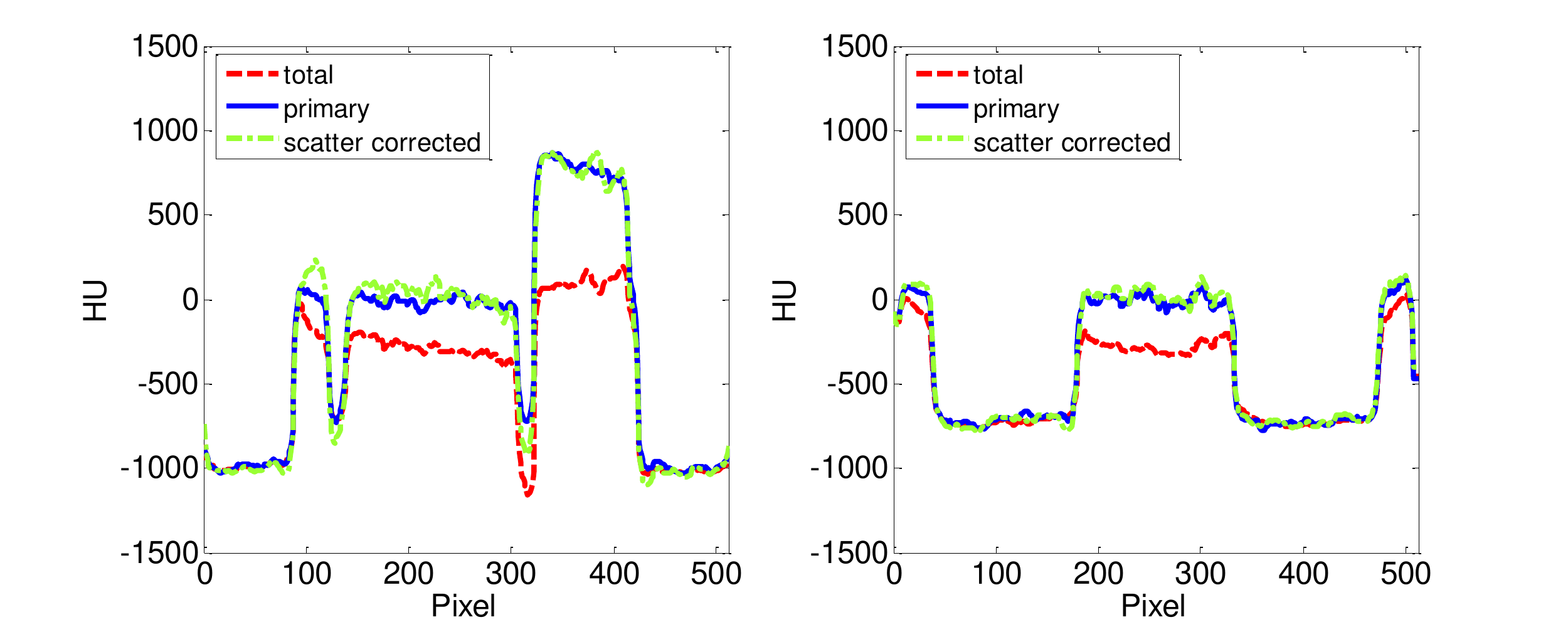}
    \caption{Line profiles of the quasi-thorax phantom with polychromatic $100~kVp$ X-ray source. Shading and reduced CT number caused by scatter radiation are clearly shown by the red line. After correction with the proposed method, shading artifacts are well removed and CT number are well recovered, and the corrected profiles match the scatter-free primary profiles well. (a) Vertical direction, (b) horizontal direction. }
    \label{fig:f5}
\end{figure}

Fig.~\ref{fig:f4} shows the results of the same phantom but with the $100~kVp$ polychromatic x-ray spectrum. As shown in the corrected image and the corresponding difference image, scatter artifacts were also significantly reduced and the CT number is comparable to the primary image. The average CT number errors are reduced from -206 HU to -23 HU for the different ROIs (shown in Figure.~\ref{fig:f2} upper-left corner). Fig.~\ref{fig:f5} shows the line profiles of the quasi-thorax phantom along central vertical direction and central horizontal direction (shown in Figure.~\ref{fig:f4} upper-left corner). We can see scatter caused shading artifacts are almost completely removed and the reduced CT number are accurately recovered.

\section{Conclusion and future work}
This work presents a new approach to correct scatter artifacts for the CBCT imaging systems. In the proposed method,  instead of using a dedicated calibration measurement or Monte Carlo simulations to calibrate the free parameters in the convolution model, we used the difference of the raw projection data and the polychromatic reprojection as our estimated rough scatter profile to calibrate the free parameters in the analytical convolution kernel.  The proposed method was evaluated using Monte Carlo simulations with both monochromatic and polychromatic X-ray sources. Results demonstrated that scatter artifacts were almost completely removed in both cases. Significant improvements have been achieved in both image uniformity and CT number accuracy. Note that the proposed correction algorithm is intrinsically patient-specific.  The proposed method avoids the need of extra hardware support, system modifications, or prolonged data acquisition time. Thus the proposed method can be potentially employed to correct scatter artifacts for existing CBCT scanners.

There are several analytical scatter estimation models. For example, scatter signal can be expressed as weighted summation of a set of sine and cosine basis function. The proposed method can be applied to these analytical scatter estimation models to calibrate the unknown weight. Further research will be focused on other scatter models.

\section*{}
\vspace{-2em}
This work has not been submitted to any other conferences or publications.
\vspace{-1em}

\bibliography{ScatterArtifactsCorrection}
\bibliographystyle{spiebib}

\end{document}